# Cage-like Au$_{32}$ detected by calculated optical spectroscopy

Wei Fa, Jian Zhou, Chuanfu Luo, and Jinming Dong*

*Group of Computational Condensed Matter Physics,*

*National Laboratory of Solid State Microstructures,*

*and Department of Physics, Nanjing University, Nanjing, 210093, China*

## Abstract

The optical absorptions of different Au$_{32}$ isomers in the whole frequency range from the far-infrared (FIR) to near-ultraviolet (UV) have been calculated using the relativistic density-functional method in order to identify their geometrical structures. It is found that there exists a distinctive difference between the absorption spectra of the icosahedral cage-like Au$_{32}$ and its amorphous isomers. The former shows significant absorption peaks in the visible and near-UV range, and a characteristic FIR-active mode at 46 cm$^{-1}$, making it possible to be distinguished unambiguously from others, which suggests that the optical spectra can thus be used as an efficient experimental tool to detect the "golden fullerene".

## I. INTRODUCTION

Gold clusters have received special attention due to their potential applications in catalysis and nanostructured material science.[1–4] For the cluster size up to several hundred atoms, its geometric structure is strongly size-dependent, which, in turn, impacts on its electronic, optical, and chemical properties. Therefore, it is fundamentally important to identify their geometric structures for the controlled use of clusters in future nanotechnology. In the past decades, the medium-sized gold clusters have been extensively studied. Some systematic numerical studies based on the empirical potentials predicted that the amorphous compact packing is dominant for $Au_N$ with $N > 14$.[5–7] Recently, J. Li *et al.* revealed the unique pyramid $Au_{20}$ structure, which is a bulk gold fragment with a small structural relaxation, deduced from the photoelectron spectroscopy and comparison with the relativistic density-functional theory (DFT) calculations.[8] More recently, a novel cage-like structure of $Au_{32}$ with $I_h$ symmetry was predicted by the relativistic DFT to be the ground state and chemically stable because of its extremely large energy gap between the highest occupied molecular orbital (HOMO) and the lowest unoccupied molecular orbital (LUMO) up to 1.7 eV.[9] This perfect $I_h$ structure, referred as "golden fullerene", can be constructed from the $C_{60}$ as a template. The discovery of cage-like $Au_{32}$ would be of practical use, for example, to accommodate other atoms or molecules.[10] Another icosahedral golden fullerene $Au_{42}$, born out of the $I_h$ $C_{80}$, was also studied numerically, which is found to be competitive in energy with its compact isomers.[11]

Despite these theoretical predictions, there is still no experimental evidence up to now for the "golden fullerene". We do not know even what kind of experimental methods could be used to detect correctly the cage-like $Au_{32}$. The previous works on the $Au_{32}$ depended on the total energy to characterize the stability of the fullerene cage.[9,10] On the other hand, less is known concerning the optical properties of different $Au_{32}$ isomers, which is disadvantageous to answer whether the golden fullerene really exists in the medium-sized range. In view of this situation, the spectroscopic tools probably provide a better choice to identify the geometric structure of a cluster, which has been successfully done for the small anionic $Au_N^-$ ($N = 7$-11).[12]

Therefore, we report here our relativistic DFT simulations on the relationship between the geometrical structures and the optical responses from the infrared (IR) to ultraviolet (UV) spectra for the different isomers of $Au_{32}$, based upon which the $I_h$ $Au_{32}$ cage can be distinguished unambiguously from others, e.g., the previously suggested space-filled structures. This paper is organized as follows. In Section II, we introduce the details of DFT computational procedure. The obtained low energy structures of $Au_{32}$ and their optical spectroscopies are discussed in Section III. Some concluding remarks are offered in Section IV.

## II. COMPUTATIONAL DETAILS

The optimized geometries and corresponding absorption spectra of gold clusters were calculated using Accelry's DFT program DMol$^3$.[13] A relativistic semi-core pseudopotential[14] and a double-numeric-polarized basis set were chosen to do the electronic structure calculations. For the exchange-correlation functional, the spin-polarized generalized gradient approximation was used.[15] A real space cutoff of 7.0 Å was used for the numerical integration. All SCF procedures were done with a convergence criterion of $10^{-6}$ a.u. on the total energy and electron density. The cluster geometry was optimized by the Broyden-Fletcher-Goldfarb-Shanno algorithm[16] without symmetry constraints until the total energy was converged to $10^{-5}$ eV in the self-consistent loop and the force on each atom was less than 5 meV/Å. As for the initial geometries, we started with not only the available structures in the literature,[5–7,9,10] but also new ones obtained by an unbiased global search with the guided simulated annealing[17] to an empirical interaction potential.[18]

Once the equilibrium geometry was obtained, the vibrational spectra were evaluated by diagonalizing the force constant matrix. The IR intensities were determined from the derivative of the electric dipole moment. The absorption spectra were calculated in the dipole approximation using the dipole transition between the ground state and excited state. The DMol$^3$ assumed that the ground state is a Slater determinant formed by the occupied Kohn-Sham orbitals, and the excited state is obtained from the ground state by eliminating one electron from an occupied orbital and placing it into an unoccupied orbital. The second assumption for the excited state is quite drastic, which ignores both relaxation effects and more fundamental problems within DFT. For example, the DFT calculations always underestimate the HOMO-LUMO gap, which may lead the absorption peaks red-shift. However, the comparison of the optical absorption between different isomers can still give the useful information of detecting their geometrical characteristics as we shall show in Section III.

The accuracy of the current computational scheme has been checked by benchmark calculations on the gold atom and the dimer. The ionization potential of 9.71 eV and the electron affinity of 2.20 eV are obtained for the gold atom, which agree with the corresponding experimental data of 9.22 eV[19] and 2.31 eV,[20] respectively. We also note that for the $Au_2$, the calculated binding energy of 2.16 eV, bond length of 2.54 Å, and vibration frequency of 173 cm$^{-1}$, are all consistent with the experimental data of 2.28 ± 0.10 eV, 2.47 Å, and 191 cm$^{-1}$, respectively,[21,22] and with previous DFT results of 2.43 eV, 2.55 Å, and 173 cm$^{-1}$ by J. Wang *et al.*[23]

## III. RESULTS AND DISCUSSIONS

Sixteen different candidates of $Au_{32}$, divided into three distinct groups, i.e., cage-like, space-filled, and bulk-fragment, have been considered in our DFT scheme. In good agreement with previous results,[9,10] the icosahedral cage is found to be the most stable structure, well separated by a significant energy gap of 0.96 eV from the first close-lying isomer with $D_{6h}$ symmetry. Though the low-symmetrical space-filled structures have higher energies than the icosahedral cage, they coexist with other hollow cage-like structures, all of which constitute a series of isomers with an almost continuous energy distribution. On the other hand, the bulk-fragment isomers not only have higher energies than the cage-like and space-filled structures, but also have some modes with "imaginary" frequencies, indicating that they are not stable. Therefore, in the following, we focus our attention to the cage-like and space-filled isomers of $Au_{32}$, among which six different lowest-energy configurations are illustrated in Fig. 1 together with their absorption spectra. The structures (a)-(c) belong to the cage-like species without core atoms, among which the structure (c) is a new low-energy candidate constructed by closing a short segment of (6, 0) gold nanotube with two rhombi, and so could be called as a tube-like gold cluster too.[24] The structures (d)-(f)[25] are space-filled or compact, among which the structure (e) was reported previously to be the "ground state" of the empirical potentials.[6]

From Fig. 1, it is especially impressive to see an evident difference between the absorption spectra of the cage-like and space-filled structures, suggesting that the optical spectroscopy can be used to sensitively detect the cage-like $Au_{32}$. Most of the absorption peaks locate in the visible and near-UV region, easily detected by further experimental measurements. The icosahedral cage (a) shows two main peaks at 1.77 and 2.43 eV, respectively, among which the first one is very sharp, but the second one has two shoulders on its both sides, lying at 2.35 and 2.58 eV, respectively. The cage (a) has also extra absorptions with much less intensities in the near-UV range. The spectrum of the $D_{6h}$ cage (b) exhibits a rich structure in the range from 1 to 5 eV, in which at least eight well-resolved peaks are observed. Finally, two main peaks centered at 1.64 and 2.94 eV are found to dominate the spectrum of the cage-like structure (c), which, however, are separated by a much broader interval of 1.3 eV than that of the icosahedral cage (only 0.66 eV). The evident difference in the absorption of the three cage-like structures seems to root in their distinct symmetries of the cage-like geometrical structures because the structure (a) is more symmetrical and compact in space than the isomers (b) and (c).

On the other hand, the space-filled structures (d)-(e) show a more continuous distribution in the calculated spectra. It is somewhat difficult to distinguish one from another based upon their absorption properties since all of these space-filled isomers show a similar feature in the visible range. There are only smaller differences between their spectra: the isomers (d) and (e) contain, respectively, double peaks at about 2.13 and 2.28 eV, and three ones at 2.15, 2.44, and 2.74 eV. And the cluster (f) with an approximate $C_{3v}$ symmetry presents a more well-defined spectrum with a sharp peak at 2.38 eV.

We would like to emphasize that the characteristics shown in the calculated $Au_{32}$ absorption spectra can be well used to identify definitely the $I_h$ $Au_{32}$ cage from others, including both of the amorphous structures and other cage-like isomers. In order to understand better the absorption properties, we have further analyzed the structural properties of the lowest energy cage-like and space-filled isomers, illustrated by the distribution of interatomic distances (DID) and shown in Fig. 2. The DID of the space-filled structure (d) shows a remarkable tendency toward more uniform distribution. The number of unequivalent atoms increases greatly from the cage (a) to the space-filled isomer (d), which determines whether the spectrum shows the well-separated peaks or the broadly continuous distribution.

Since the vibrational spectroscopy depends sensitively on the geometrical structure, we have calculated their vibrational density of states (VDOS) and corresponding far-infrared (FIR) spectra, shown also in Fig. 2. The vibration spectra of the $I_h$ $Au_{32}$ are in the frequency range of 30.4 - 50.8 cm$^{-1}$, which agrees well with the range of 30 - 145 cm$^{-1}$ obtained by the BP86 functional and the range of 37 - 147 cm$^{-1}$ got with the hybrid PBE0 functional in Ref. 9. However, the lowest vibrational frequencies of its space-filled isomers are shifted to lower frequencies, making them less stiff than the $I_h$ Au$_{32}$. For example, the frequency range of the lowest-energy space-filled isomer is 5.9 - 173.0 cm$^{-1}$. As expected, several distinct peaks are well resolved in the VDOS of the icosahedral Au$_{32}$ cage, and more importantly, its FIR spectrum exhibits a strong peak at 46 cm$^{-1}$ with an intensity of 4.6 km/mol, which could be used as a fingerprint signal to detect the icosahedral Au$_{32}$. It has also two much weaker peaks at 96 and 125 cm$^{-1}$, respectively, due to the slight structural distortion from the perfect icosahedron. On the other hand, almost all the modes of the space-filled structure (d) are infrared-active, showing a spectrum with many unresolved peaks except a double-peak at around 70 cm$^{-1}$, indicative of its absence of spatial symmetry. However, their intensities are much lower than the strong absorption of the cage (a). Significant differences of the FIR absorption spectra between the cage-like and space-filled structures provide another clue to verify the existence of "golden fullerene". We hope that the infrared resonance-enhanced multiple photon dissociation can be used to acquire the FIR absorption spectra of Au$_{32}$, as those done for vanadium and niobium clusters.[26–28] In addition, we also notice that the $I_h$ Au$_{32}$ cage has a Raman-active breathing mode at 91 cm$^{-1}$, characteristic of the cage or tube-like structures, which is similar to those found in carbon fullerenes, and so may be tested by future Raman spectra.

The absorption difference between the cage-like and space-filled Au$_{32}$ would also exist in other sizes of Au$_N$, such as the Au$_{42}$. In Fig. 3 (a), two well-separated main peaks, both of which exhibit the peak splitting, can be resolved for the icosahedral Au$_{42}$ cage. Also, the spectrum shows additional weak absorptions in the near-UV region around 3.21, 3.75, and 4.09 eV, etc. The whole spectrum of the cage-like Au$_{42}$ is similar to that of the icosahedral Au$_{32}$, and the difference may be due to the diameter increase from the Au$_{32}$ to the Au$_{42}$ cage. On the other hand, as shown in the inset of Fig. 3 (b), the amorphous features appear in the DID of the space-filled Au$_{42}$ isomer with $C_s$ symmetry, which was reported as the "global minimum" based on the Sutton-Chen potentials with. Compared with the high-symmetrical $I_h$ Au$_{42}$ cage, the large numbers of unequivalent gold atoms in this space-filled Au$_{42}$ structure induce the broad absorptions peak centered at about 2.1 eV, seen clearly from Fig. 3 (b), which is obviously different from the well-defined spectrum obtained for the $I_h$ Au$_{42}$ cage shown in Fig. 3 (a).

## IV. SUMMARY

We have shown the structure-dependent optical absorption of $Au_{32}$ employing the DFT techniques. The peak positions and the overall spectra exhibit distinctive features between its cage-like and amorphous isomers, not only in the visible and near-UV, but also in the FIR frequency range. Especially, the high-symmetrical icosahedral $Au_{32}$ can be identified unambiguously from its other isomers by its two strong absorptions at 1.77 and 2.43 eV, and also by its unique characteristic FIR-active mode at 46 cm$^{-1}$. Our results might shed light on how the cage-like $Au_N$ can be practically detected by their optical responses.

## ACKNOWLEDGMENTS

We acknowledge discussions with Prof. X. G. Gong. This work was supported by the Natural Science Foundation of China under grant No. A040108 and the state key program of China through grant No. 2004CB619004. The DFT calculations were made on the SGI Origin-3800 and 2000 supercomputers.
.

---

[*] Corresponding author E-mail:jdong@nju.edu.cn

[1] B. Yoon, H. Häkkiinen, U. Landman, A. S. Worz, J.-M. Antonietti, S. Abbet, K. Judai, and U. Heiz, Science **307**, 403 (2005).

[2] P. Schwerdtfeger, Angew. Chem. Int. Ed. **42**, 1892 (2003).

[3] M.-C. Daniel and D. Astruc, Chem. Rev. **104**, 293 (2004).

[4] S. Chen, R. S. Ingram, M. J. Hostetler, J. J. Pietron, R. W. Murray, T. G. Schaaff, J. T. Khoury, M. M. Alvarez, and R. L. Whetten, Science **280**, 2098 (1998).

[5] J. P. K. Doye and D. J. Wales, New J. Chem. **22**, 733 (1998).

[6] N. T. Wilson and R. L. Johnston, Eur. Phys. J. D **12**, 161 (2000).

[7] S. Darby, T. V. Motimer-Jones, R. L. Johnston, and C. Roberts, J. Chem. Phys. **116**, 1536 (2002).

[8] J. Li, X. Li, H. Zhai, and L.-S. Wang, Science **299**, 864 (2003).

[9] M. P. Johansson, D. Sundholm, and J. Vaara, Angew. Chem. Int. Ed. **43**, 2678 (2004).

[10] X. Gu, M. Ji, S. H. Wei, and X. G. Gong, Phys. Rev. B **70**, 205401 (2004).

[11] Y. Gao and X. C. Zeng, J. Am. Chem. Soc. **127**, 3698 (2005).

[12] S. Gilb, K. Jacobsen, D. Schooss, F. Furche, R. Ahlrichs, and M. M. Kappes, J. Chem. Phys. **121**, 4619 (2004).

[13] B. Delley, J. Chem. Phys. **92**, 508 (1990).

[14] B. Delley, Phys. Rev. B **66**, 155125 (2002).

[15] J. P. Perdew, in *Electronic Structure of Solids '91*, edited by P. Ziesche and H. Eschrig (Akademie Verlag, Berlin, 1991); J. P. Perdew, J. A. Chevary, S. H. Vosko, K. A. Jackson, M. R. Pederson, D. J. Singh, and C. Fiolhais, Phys. Rev. B **46**, 6671 (1992).

[16] R. Fletcher, *Practical Methods of Optimization* (Wiley, New York, 1980), Vol. 1.

[17] C. I. Chou and T. K. Lee, Acta Crystallogr., Sect. A: Found. Crystallogr. A **58**, 42 (2001).

[18] F. Cleri and V. Rosato, Phys. Rev. B **48**, 22 (1993).

[19] C. Kittle, *Introduction to Solid State Physics*, 7th ed. (Wiley, New York, 1996).

[20] J. E. Huheey, E. A. Keiter, and R. L. Keiter, *Inorganic Chemistry: Principles of Structure and Reactivity*, 4th Ed. (HarperCollins, New York, 1993).

[21] *CRC Handbook of Chemistry and Physics*, 55thed., edited by R. C. Weast (CRC Press, Cleveland, OH, 1974).

[22] *American Institute of Physics Handbook* (McGraw-Hill, New York, 1972).

[23] J. Wang, G. Wang, and J. Zhao, Phys. Rev. B **66**, 35418 (2002).

[24] Y. Oshima, A. Onga, and K. Takayanagi, Phys. Rev. Lett. **91**, 205503 (2003).

[25] The initial structure for the cluster (f) is the ground state of the empirical potentials (see Ref. 5, 7), which, after the further DFT optimization, converts into a more symmetrical $C_{3v}$ structure, being a fragment of the hexagonal closest packing with a structural relaxation.

[26] A. Fielicke, A. Kirilyuk, C. Ratsch, J. Behler, M. Scheffler, G. von Helden, and G. Meijer, Phys. Rev. Lett. **93**, 023401 (2004).

[27] C. Ratsch, A. Fielicke, A. Kirilyuk, J. Behler, G. Helden, G. Meijer, and M. Scheffler, J. Chem. Phys. **122**, 124302 (2005).

[28] A. Fielicke, C. Ratsch, G. Helden, and G. Meijer, J. Chem. Phys. **122**, 091105 (2005).

**Figure Captions**

FIG. 1. (Color online) Calculated optical absorption spectra of the cage-like (left) and space-filled (right) $Au_{32}$ isomers. The corresponding geometries and the relative energies to the icosahedral cage are also shown in each panel.

FIG. 2. (Color online) DID, VDOS, and the calculated FIR absorption spectra for the lowest energy cage-like (a) (left panel) and space-filled (d) (right panel) $Au_{32}$ indicated in Fig. 1. The shaded areas describe the IR spectra.

FIG. 3. (Color online) Calculated optical absorption spectra for two different $Au_{42}$ isomers: (a) the icosahedral cage and (b) the space-filled structure. The corresponding configurations are drawn in each panel. Insets are the distribution of interatomic distances for them.

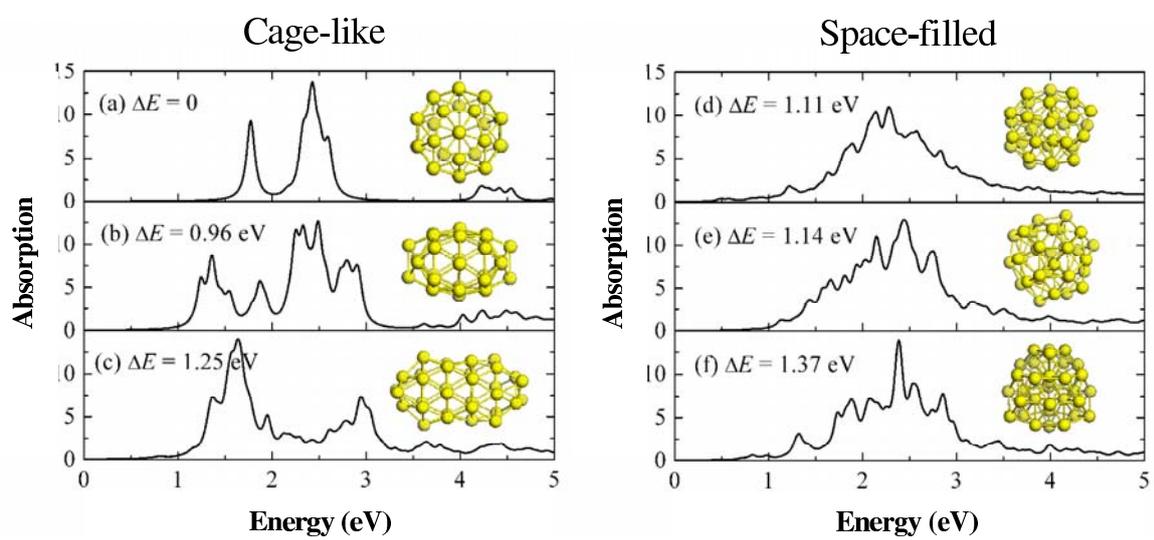

Figure 1

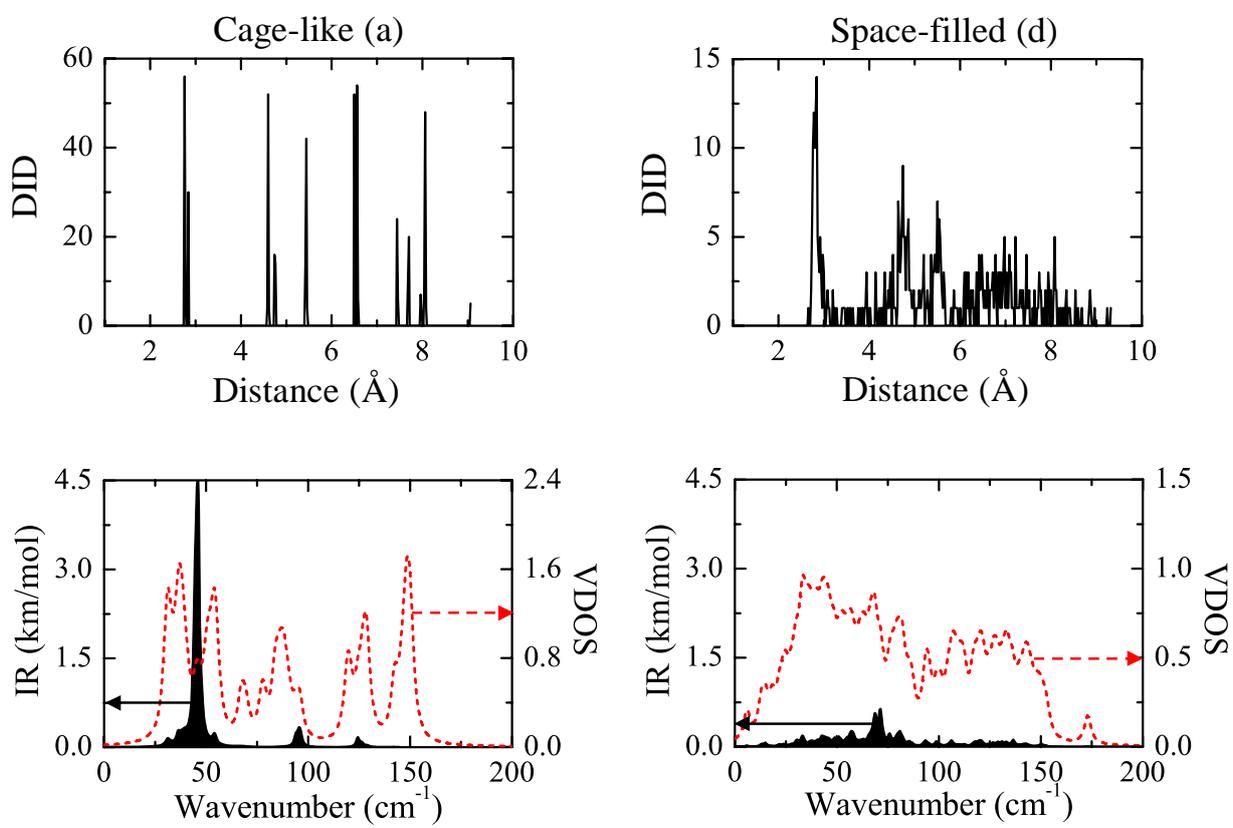

Figure 2

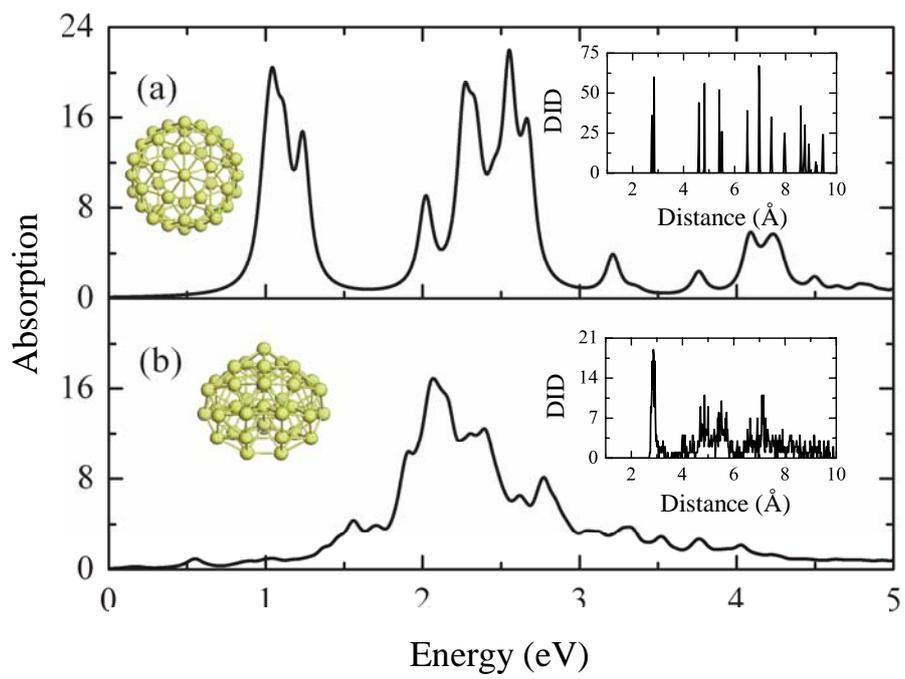

Figure 3